\documentclass{article}
\usepackage[utf8]{inputenc}
\usepackage[margin=0.8in,tmargin=0.5in,bmargin=1in]{geometry}
\usepackage{parskip}
\usepackage{amsmath}
\usepackage{amssymb}
\usepackage{mathrsfs}
\usepackage{braket}
\usepackage{graphicx}
\usepackage{caption}
\usepackage{subcaption}
\usepackage{array}
\usepackage{authblk}
\usepackage{float}
\usepackage{hyperref}
\hypersetup{colorlinks=true, linkcolor=blue, citecolor=blue, filecolor=magenta, urlcolor=blue}

\usepackage[backend=biber,
    style=numeric-comp,
    sorting=none,
    firstinits=true,
    maxnames=7,
    date=year,
    isbn=false,
    doi=false,
    url=false]{biblatex}

\AtEveryBibitem{%
  \clearfield{issue}%
  \clearfield{number}}

\renewbibmacro{in:}{}

\DeclareFieldFormat[article]{volume}{\mkbibbold{#1}}

\newbibmacro{string+doi}[1]{%
  \iffieldundef{doi}{#1}{\href{http://dx.doi.org/\thefield{doi}}{#1}}}

\DeclareFieldFormat{title}{\usebibmacro{string+doi}{\mkbibemph{#1}}}

\DeclareFieldFormat[article]{title}{\usebibmacro{string+doi}{\mkbibquote{#1}}}

\DefineBibliographyStrings{english}{%
  page             = {},
  pages            = {},
}

\allowdisplaybreaks

\newcommand{\angleaverage}[1]{\ensuremath{\left\langle #1 \right\rangle}}

\ExplSyntaxOn
\NewDocumentCommand \colVec { m }
  {
    \begin{pmatrix}
      \clist_use:nn {#1} { \\ }
    \end{pmatrix}
  }
\ExplSyntaxOff

\newcommand{\figRef}[2]{\ref{#1}\hyperref[#1]{#2}}

\title{\huge{Pressure and Proximity Tuned Twisted Bilayer Graphene}}
\author{David T. S. Perkins}
\author{Joseph J. Betouras}
\affil{\textit{Department of Physics, Loughborough University, Loughborough LE11 3TU, England, United Kingdom}}
\date{}

\begin{document}

\maketitle

\begin{abstract}
    The coupling between layered atomically-thin materials mediated by van der Waals forces allows strong electronic correlations, unique topological signatures, and non-trivial spin textures, all of which play an important role in the creation of spin-, valley-, and orbitronic devices. Here, we demonstrate that twisted bilayer graphene encapsulated by transition metal dichalcogenides exhibits a generically non-radial spin texture yet hosts a purely collinear Edelstein effect despite this lack of radial symmetry. Moreover, we show how uniaxial pressure can be used to further tune the band structure and change the number of active Fermi surfaces without compromising the collinear response. Lastly, we illustrate how the quantum geometry changes in the encapsulated twisted graphene bilayer with larger pressures spreading the Berry curvature over large regions of the moir\'{e} Brillouin zone. These results illustrate how encapsulation and pressure can be used to drastically alter the topology and spin-charge interconversion processes of moir\'{e} heterostructures.
\end{abstract}

\section{Introduction}

Since the synthesis of the first truly atomically-thin material, graphene \cite{Novoselov2004}, a plethora of two-dimensional (2D) materials have attracted much attention on both experimental and theoretical fronts in the study of strong correlations, quantum geometry, and transport. It is natural to stack these 2D materials in a layer-by-layer fashion to explore how proximity coupling due to weak van der Waals (vdW) forces can alter the energetic landscape of the electrons and in turn allow us to engineer emergent properties of such systems. Common materials considered in these studies are graphene \cite{CastroNeto2009,McCann2013,Moon2019,Yu2019,Zhou2021_SC,Zhou2021_qtrMetal,Sousa2022,Pan2025,Cuypers2026}, hexagonal boron nitride \cite{Drogler2014,Jung2015,Huang2022,Durand2023}, and transition metal dichalcogenides (TMDs) \cite{Eda2012,Wang2012,Mak2014,Qian2014,Fang2015,Voiry2015,Mak2016,Tang2017,Wang2018,Wu2018,Shi2019}. The aim of combining these 2D materials to create bespoke devices is vast and include enhancing spin-charge interconversion \cite{Avsar2020,Sierra2021,Perkins2023}, enabling superconductivity \cite{Cao2018,Zhou2021_SC,Su2023,Xia2026}, accessing the orbital degrees of freedom \cite{Pezo2023,Veneri2025,Cysne2026}, generating non-Fermi liquid behaviour \cite{Cao2021,Jaoui2022,Wei2024,Xia2026}, creating quantum dots and qubits \cite{Gachter2022,Perkins2024,Dulisch2025,Gerber2025}, exerting torque on magnetic alignment \cite{Go2020,Sousa2020,Veneri2022}, and altering quantum geometry \cite{Liu2024,Pan2025,Yu2025,Cuypers2026}.

In recent years, driven by the observation of unconventional superconductivity in magic-angle twisted bilayer graphene (TBG) \cite{Cao2018}, rotational misalignment of the stacked layers (i.e. twisting) has taken centre stage in our endeavour to further tune vdW heterostructures. This new field of twistronics is a multifaceted theme that overlaps significantly with several other fields because the systems display strong correlations and non-trivial topology. Many of these works have focused on the emergence of unconventional superconductivity \cite{Su2023,Xia2026}, strange metallicity \cite{Cao2021,Jaoui2022,Wei2024,Xia2026}, and the interplay of topology and Wannierisation \cite{Koshino2018,Song2022,Ledwith2025,Hu2026arxiv,Vituri2026arxiv,Wie2026arxiv} in honeycomb-based vdW heterostructures, whilst recent works have begun to generalise the study of moir\'{e} systems to other lattices including the dice and Lieb lattices \cite{Zhou2024} and the kagome lattice \cite{Lima2019,Perkins2025b,Perkins2026arxiv,Hung2026arxiv}. However, there have been comparatively few studies regarding the potential of these twist-engineered proximitised systems in the absence of strong correlations. Naturally, graphene-based vdW heterostructures have once again found themselves at the forefront of studies into how twisting can alter proximity-induced couplings. In particular, the Hamiltonian for a graphene monolayer stacked on top of a TMD layer was shown to change substantially when the two layers were rotationally misaligned, with the Rashba spin-orbit coupling (SOC) term acquiring a twist-dependent phase -- the \textit{Rashba phase} -- that rotated the momentum-space spin texture of the Fermi surface, $\langle \mathbf{S}_{\mathbf{k}} (\theta) \rangle_{\text{FS}}$ \cite{Li2019}. Moreover, the magnitude of the proximity-induced Rashba and spin-valley SOCs were predicted \cite{David2019,Peterfalvi2022} and demonstrated \cite{Rao2023,Sun2023} to vary drastically with twist angle, with giant suppression of the Rashba SOC from its maximal magnitude and the complete vanishing of the spin-valley SOC for a $30^{\circ}$ twist. Through the introduction of the Rashba phase, it was shown that there may exist a critical angle, $\theta = \theta_{c}$, that yields a purely radial spin texture, $\hat{\mathbf{k}} \cdot \langle \mathbf{S}_{\mathbf{k}} (\theta_{c}) \rangle_{\text{FS}} = |\langle \mathbf{S}_{\mathbf{k}} (\theta_{c}) \rangle_{\text{FS}}|$, thus replacing the conventional Edelstein effect (EE) -- where an electrical current generates a non-equilibrium perpendicular in-plane spin polarisation ($\mathbf{J} \cdot \mathbf{S} = 0$) -- with a collinear Edelstein effect (CEE; $|\mathbf{J} \cdot \mathbf{S}| = |\mathbf{J}| |\mathbf{S}|$) \cite{Veneri2022}. Moreover, a microscopic theory of the spin Hall effect -- the generation of a spin current perpendicular to an applied electrical current -- in proximitised graphene accounting for disorder effects in both the perturbative weak scattering and non-perturbative unitary limits was only recently developed \cite{Perkins2024}, illustrating giant disorder denormalisation and how the spin Hall effect (SHE) can be switched off entirely or enhanced through twisting.

A crucial next step is to consider the possible spin textures that may manifest in TBG, as these are expected to influence the processes that involve both spins and charges. In Ref. \cite{Tan2024}  it was shown that magic angle TBG (MATBG) with one layer proximitised by a TMD hosts three topological phases characterised by the valley Chern numbers of the spin split conduction and valence bands and skyrmion-like spin textures around $\Gamma_{\text{M}}$ which could be tuned using out-of-plane electric fields. However, as we demonstrate below, these spin textures yield an Edelstein response that is a mixture of the conventional and collinear effects.

In this paper, we consider MATBG encapsulated by two TMDs and demonstrate that the emergent spin texture yields only the CEE and SHE. We further show that uniaxial out-of-plane pressure can be used to enhance the tunnelling between the graphene layers to alter the band structure, generating new band inversions and thus changing the topological phase. This out-of-plane pressure influences the concentration of Berry curvature causing it to spread out across larger portions of the moir\'{e} Brillouin zone (MBZ). Moreover, pressure tuning changes the number of bands crossing the Fermi energy, and hence the number of active spin textures. The structure of this paper is as follows: Section \ref{Sec_model} introduces the Bistritzer-MacDonald formalism for constructing the low-energy continuum Hamiltonian for TBG and how we apply it to the encapsulated scenario. This section also presents how we model pressure in the TBG system by way of reducing the interlayer separation. We then present the band structures and spin textures of the encapsulated MATBG system for different choices of the proximity-induced SOC in Section \ref{Sec_ES_ST}. Following, in Section \ref{Sec_QG}, we demonstrate how the Berry curvature changes under pressure. Finally, we summarise and discuss our results in Section \ref{Sec_conclusions}.

\section{Model} \label{Sec_model}

\subsection{Hamiltonian}

The Bistritzer-MacDonald continuum Hamiltonian for TBG at small twist angles is essentially constructed from two fundamental ingredients: the individual layer Hamiltonians describing the separate proximitised graphene monolayers, $H_{\mathbf{k}}^{\tau(l)}$, and the interlayer tunnelling Hamiltonian connecting the two layers, $H_{\perp,\mathbf{k}\mathbf{p}}^{\tau}$. For a detailed derivation of these types of Hamiltonians, we direct the reader to the literature outlining these details \cite{Lopes_dos_Santos2007,Bistritzer2011,Lopes_dos_Santos2012,Catarina2019,Koshino2019,Song2019,Bernevig2021,Perkins2026arxiv}. Here we use $\tau = \pm 1$ to denote the monolayer valley index corresponding to the monolayer Brillouin zone corner at $\mathbf{K}_{\tau} = \tau \frac{4\pi}{3a} (1,0)$ ($a$ is the lattice constant), and $l =1,2$ to label the layers. The full low-energy Hamiltonian for the TBG system, with a twist of $\theta$ and both layers being proximitised, may then be written as ($\hbar = 1$)
\begin{subequations}
\begin{gather}
    \mathcal{H}_{\mathbf{k}}^{\tau} = \begin{pmatrix}
        H_{\mathbf{k}}^{\tau(1)} & H_{\perp,\mathbf{k}\mathbf{p}}^{\tau} \\
        H_{\perp,\mathbf{k}\mathbf{p}}^{\tau\dagger} & H_{\mathbf{p}}^{\tau(2)}
    \end{pmatrix},
    \qquad
    H_{\perp,\mathbf{k}\mathbf{p}}^{\tau} = \sum_{i} T_{i}^{\tau} \delta_{\mathbf{p}-\mathbf{k},\tau \mathbf{q}_{i}}^{\null},
    \\
    T_{j}^{\tau} = w_{0} \gamma_{00} + w_{1} \cos\left( \eta_{j} \frac{2\pi}{3}\right) \gamma_{x0} + w_{2} \sin\left( \tau \eta_{j} \frac{2\pi}{3}\right) \gamma_{y0}, \qquad \eta_{j} = \delta_{j,tr} - \delta_{j,tl}
    \\
    \mathbf{q}_{b}^{\null} = \frac{8\pi}{3a} \sin\left(\frac{\theta}{2}\right) \colVec{0 \\ -1},
    \qquad
    \mathbf{q}_{tr}^{\null} = \frac{4\pi}{3a} \sin\left(\frac{\theta}{2}\right) \colVec{\sqrt{3} \\ 1},
    \qquad
    \mathbf{q}_{tl}^{\null} = \frac{4\pi}{3a} \sin\left(\frac{\theta}{2}\right) \colVec{-\sqrt{3} \\ 1},
    \\
    \begin{split}
        H_{\mathbf{k}}^{\tau(l)} = v(\tau \gamma_{x0}^{\null} k_{x}^{\null} + \gamma_{y0}^{\null} k_{y}^{\null}) &+ \lambda_{\text{sv}}^{(l)}(\tilde{\theta}_{l}^{\null}) \tau \gamma_{0z} + (\delta_{1,l}^{\null} - \delta_{2,l}^{\null}) \frac{U}{2} \gamma_{00}^{\null}
        \\
        &+\lambda_{\text{R}}^{(l)}(\tilde{\theta}_{l}^{\null}) \exp\left(i\frac{\alpha_{\text{R}}^{(l)}(\tilde{\theta}_{l}^{\null})}{2} \gamma_{0z}^{\null} \right) (\gamma_{xy}^{\null} - \gamma_{yx}) \exp\left(-i\frac{\alpha_{\text{R}}^{(l)}(\tilde{\theta}_{l}^{\null})}{2} \gamma_{0z}^{\null} \right),
    \end{split}
\end{gather}
\label{Hamiltonian}%
\end{subequations}
where $v$ is the Fermi velocity of the Dirac electrons, $\gamma_{ij} = \sigma_{i} \otimes s_{j}$ with $\{\sigma_{i}\}$ ($\{s_{i}\}$) being the set of Pauli matrices supplemented with identity acting on the sublattice (spin) degree of freedom, $\lambda_{\text{R}}^{(l)}$ ($\lambda_{\text{sv}}^{(l)}$) is the proximity-induced Rashba (spin-valley) SOC in layer $l$, $\alpha_{\text{R}}^{(l)}(\tilde{\theta}_{l}^{\null})$ is the Rashba phase induced by the neighbouring TMD at a relative angle $\tilde{\theta}_{l}^{\null}$ in layer $l$, $\delta_{i,j}$ is the Kronecker delta, and $w_{0}$ and $w_{1}$ are the interlayer tunnelling energies characterising $AA'$ and $AB'$ hopping, respectively, upon accounting for variation in the interlayer distance due to stacking order changes in the moir\'{e} unit cell. We provide a schematic of this system in Fig. \figRef{Schematic}{a} to visualise the three twist angles present here. The parameters we use throughout this work are based on those of Koshino et al. \cite{Koshino2019}. Specifically, we take $\hbar v/a = 2.1354$ eV and $a = 0.246$ nm. The interlayer tunnelling energies and distances are discussed in more detail below as these are sensitive to the applied out-of-plane pressure.

In the rest of the paper, we focus on the first magic-angle of unproximitsed TBG, $\theta = 1.05^{\circ}$ and truncate the Hamiltonian in Eq. \ref{Hamiltonian} to 10 shells (664 band Hamiltonian). We pick these numbers so that the energies of the four bands closest to charge neutrality have converged sufficiently to obey the crystal and magnetic symmetries of the moir\'{e} heterostructure.

\subsection{Interlayer Distance, Tunnelling Energies, and Pressure}

\begin{figure}
    \centering
    \includegraphics[width=\linewidth]{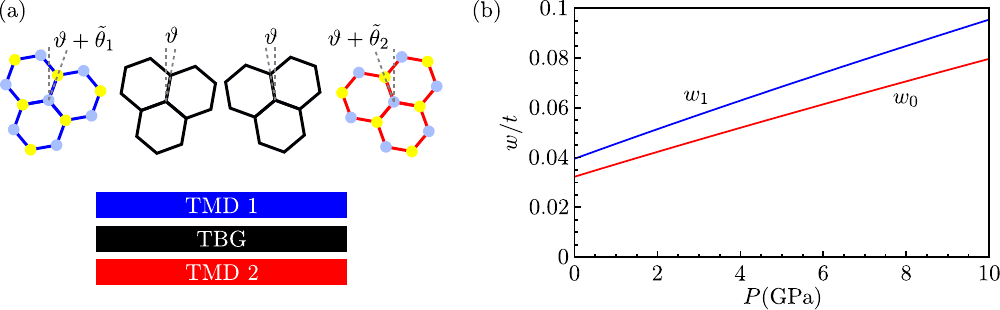}
    \caption{(a): Schematic of an encapsulated TBG sample with the twist angle between the graphene layers given by $\theta = 2\vartheta$, whilst the twists between a graphene layer and its proximitising TMD partner are denotedby $\tilde{\theta}_{l}$ ($l = 1,2$). (b): Variation of $w_{0}$ and $w_{1}$ with applied pressure assuming $d_{\perp}^{(AA)}/d_{\perp}^{(AB)}$ remains constant.}
    \label{Schematic}
\end{figure}

The interlayer distance in TBG varies throughout the moir\'{e} unit cell due to the different stacking orders at different points within the unit cell \cite{Koshino2019}. When graphene is in an $AA$ stacked region, the interlayer distance is given by $d_{\perp}^{(AA)} = 0.36$ nm, whilst $AB$ stacked regions possess $d_{\perp}^{(AB)} = 0.335$ nm. As detailed by Koshino et al. \cite{Koshino2019}, by interpolating between these two distances, we can write down a simple expression to capture how the interlayer distance changes as we displace one layer relative to the other to capture all possible stackings and not just the $D_{6h}$ and $D_{3d}$ high-symmetry stackings,
\begin{equation}
\begin{gathered}
    d_{\perp}^{\null}(\mathbf{r}) = \frac{1}{3} (d_{\perp}^{(AA)} + 2 d_{\perp}^{(AB)}) + \frac{2}{9} (d_{\perp}^{(AA)} - d_{\perp}^{(AB)}) \sum_{i=1}^{3} \cos(\mathbf{b}_{i}^{\null} \cdot \mathbf{r}),
    \\
    \mathbf{b}_{1}^{\null} = \frac{2\pi}{\sqrt{3}a} \colVec{\sqrt{3} \\ -1},
    \qquad
    \mathbf{b}_{2}^{\null} = \frac{4\pi}{\sqrt{3}a} \colVec{0 \\ 1},
    \qquad
    \mathbf{b}_{3}^{\null} = -(\mathbf{b}_{1}^{\null} + \mathbf{b}_{2}^{\null}).
\end{gathered} \label{Interlayer_interpolation}
\end{equation}

The interlayer tunnelling energy scales are sensitive to the details of the tunnelling model and thus the interlayer distance. We employ the same Slater-Koster parameterisation of the electron overlap integral from Ref. \cite{Koshino2019} ($a_{cc} = a/\sqrt{3}$ is the carbon-carbon bond length),
\begin{equation}
    - t(\mathbf{r}) = V_{pp\pi}^{0} e^{-(r-a_{cc})/(0.184 a)} \left(1 - \frac{z^{2}}{r^{2}} \right) + V_{pp\sigma}^{0} e^{-(r-d_{\perp}^{(AB)})/(0.184 a)} \frac{z^{2}}{r^{2}},
    \label{Overlap_integral}
\end{equation}
with $V_{pp\pi}^{0} = -2.7$ eV and $V_{pp\sigma}^{0} = 0.48$ eV. We then use Eqs. (\ref{Interlayer_interpolation}) and (\ref{Overlap_integral}) to determine the energy scales appearing in $H_{\perp,\mathbf{k}\mathbf{p}}^{\tau}$,
\begin{equation}
\begin{gathered}
    w_{0} = - \frac{2}{\sqrt{3} a^{2}} \int d^{2}r t(\mathbf{r} + d_{\perp}(\mathbf{r}) \mathbf{e}_{z}) e^{-i \mathbf{K}_{\tau} \cdot \mathbf{r}},
    \qquad
    w_{1} = - \frac{2}{\sqrt{3} a^{2}} \int d^{2}r t(\mathbf{r} + d_{\perp}(\mathbf{r} - \boldsymbol{\delta}_{AB}) \mathbf{e}_{z}) e^{-i \mathbf{K}_{\tau} \cdot \mathbf{r}}.
\end{gathered}
\end{equation}
In the absence of pressure, these energy scales are $w_{0} \simeq 79.7$ meV and $w_{1} \simeq 97.5$ meV \cite{Koshino2019}. The role of out-of-plane pressure is to reduce the interlayer distance, thereby enhancing the interlayer coupling and thus increasing $w_{0}$ and $w_{1}$. In TBG, the pressure-induced reduction in $d_{\perp}$ is determined by a suppression factor \cite{Wang2024},
\begin{equation}
    \alpha(P) = \left[1 -  \frac{21}{200} \ln\left(1+\frac{7}{40} \frac{P}{\text{GPa}}\right)\right].
    \label{Pressure_suppression_factor}
\end{equation}
Combining Eqs. (\ref{Interlayer_interpolation}) and (\ref{Pressure_suppression_factor}) gives $d_{\perp}(\mathbf{r},P) = \alpha(P) d_{\perp}(\mathbf{r})$ as the full function we consider to capture the effects of pressure and natural variation in layer separation due to non-uniform stacking order throughout the moir\'{e} unit cell. In constructing this function, we have assumed that the ratio $d_{\perp}^{(AA)}/d_{\perp}^{(AB)}$ is independent of the applied pressure. For illustrative purposes, we plot the variation of $w_{0}$ and $w_{1}$ in Fig. \figRef{Schematic}{b}.

\begin{figure}
    \centering
    \includegraphics[width=\linewidth]{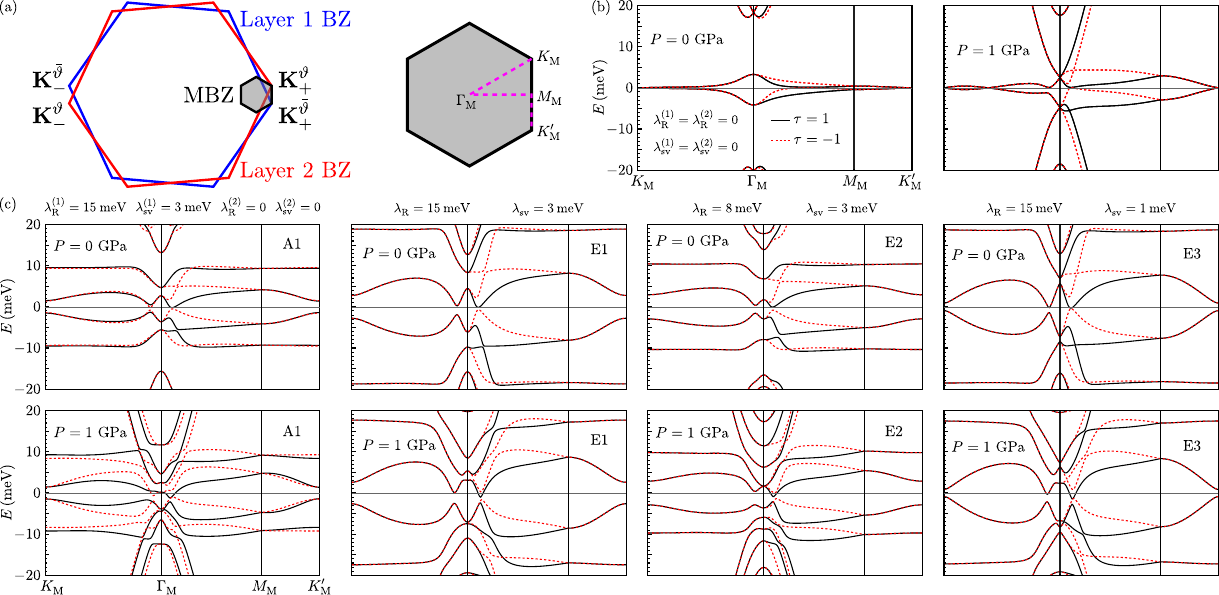}
    \caption{MBZ construction and band structures for TBG at the first magic angle of the unproximitised system, $\theta = 1.05^{\circ}$, for various choices of the proximity-induced SOCs at zero pressure and 1 GPa and a schematic for the different proximitisation setups. (a): construction of the MBZ from the BZs of the individual graphene layers. (b): Band structure of regular MATBG with no proximity induced SOC. Solid black (dashed red) lines are calculated in the $\tau = 1$ ($\tau = -1$) valley of the monolayer. (c): Band structures for proximity-coupled MATBG systems A1, E1, E2, and E3 from left to right for zero pressure (top) and 1 GPa pressure (bottom).}
    \label{Band_structures_simple}
\end{figure}

\section{Electronic Structure and Spin Texture} \label{Sec_ES_ST}

For reference, we plot the band structure for TBG without any proximity-induced SOC at zero pressure and 1 GPa in Fig. \figRef{Band_structures_simple}{a}, where we see that pressure acts to close the gap between the flat and dispersive bands. Adding SOC naturally leads to spin splitting of the flat bands, to give four bands that can be identified by their spin helicities and out-of-plane spin tilting due to Rashba and spin-valley coupling, repsectively, see Figs. \figRef{Band_structures_simple}{b}. For simplicity, we shall assume TMD encapsulation is symmetric and aligned: $\tilde{\theta}_{1,2} = 0$, which guarantees that $\alpha_{\text{R}}^{(l)} = 0$, $\lambda_{\text{R}}^{(1)} = -\lambda_{\text{R}}^{(2)} = \lambda_{\text{R}}^{\null}$, and $\lambda_{\text{sv}}^{(1)} = \lambda_{\text{sv}}^{(2)} = \lambda_{\text{sv}}^{\null}$. At a general level, we shall consider one system where only a single layer is proximitised with $(\lambda_{\text{sv}}^{(1)},\lambda_{\text{R}}^{(1)}) = (15,3)$ meV, which we shall refer to as system A1, and three encapsulated systems with differing choices of SOC denoted by E1, E2, and E3 with $(\lambda_{\text{R}}^{\null},\lambda_{\text{sv}}) = (15,3)$ meV, $(8,3)$ meV and $(15,1)$ meV, respectively; see Fig. \figRef{Band_structures_simple}{c} for sketches of these structures. We will then study systems A1 and E1 in greater depth to infer the effect of symmetric encapsulation.

\begin{figure}[H]
    \centering
    \includegraphics[width=\linewidth]{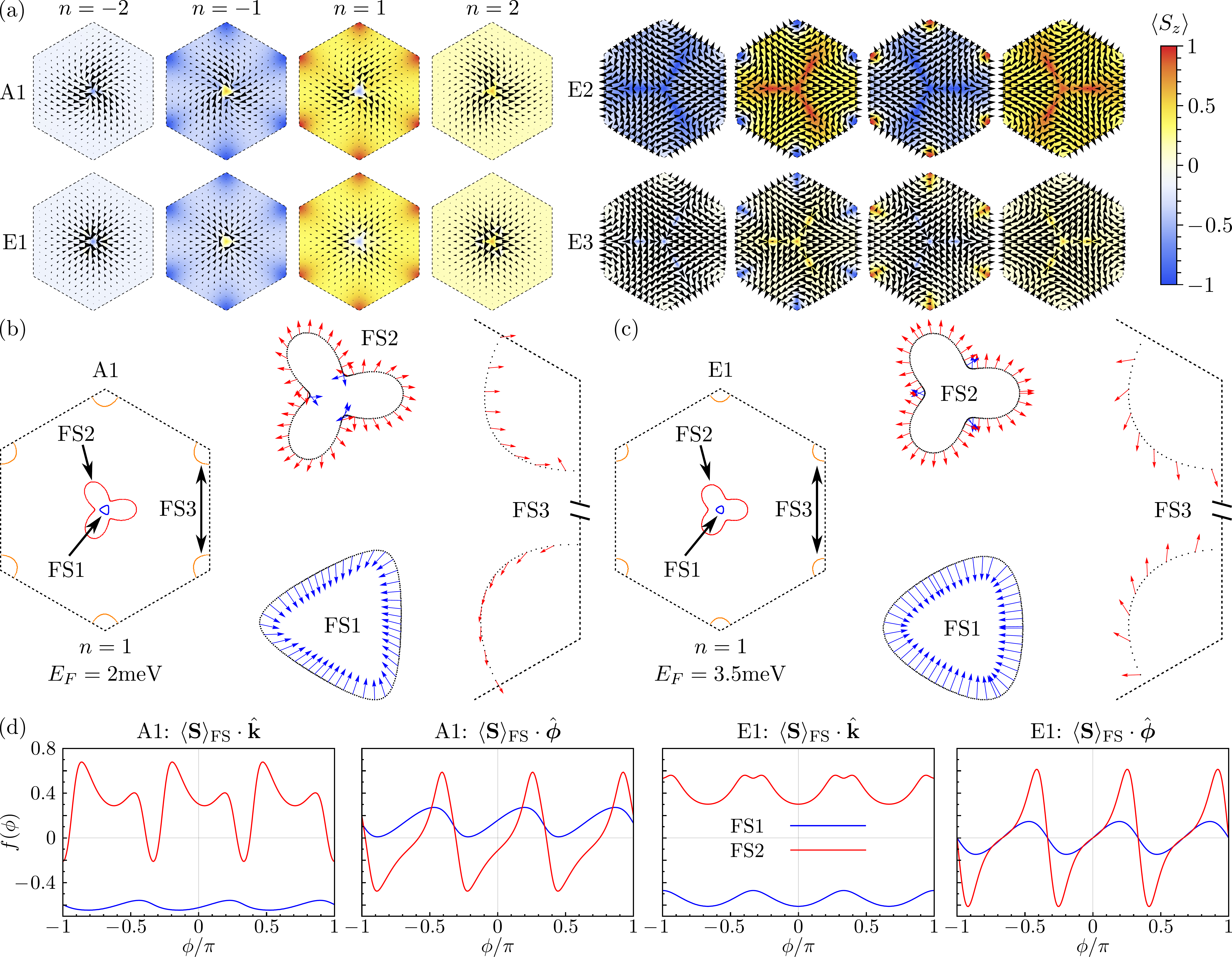}
    \caption{Zero pressure spin maps, Fermi surfaces, and spin textures. (a): Spin maps for the $|n| = 1,2$ bands of the A1, E1, E2, and E3 systems. The arrows indicate the in-plane spin projection at that point in the MBZ, with the arrow size indicating the magnitude of in-plane spin projection: larger arrow heads represent larger values of $\langle \mathbf{S}_{\parallel} \rangle_{\mathbf{k}}$. The colour density plot in the background encodes the out-of-plane spin projection. Band inversions can be seen to occur near $\Gamma_{\text{M}}^{\null}$ ($K_{\text{M}}^{\null}/K_{\text{M}}'$) for A1 and E1 (E2 and E3) in the $|n| = 1$ bands where the sign of $\langle S_{z} \rangle_{\mathbf{k}}$ can be seen to flip relative to the rest of the band. (b)-(c): Fermi surfaces and spin textures for the $n = 1$ band of the A1 system at $E_{F} = 2$ meV (b) and E1 system at $E_{F} = 3.5$ meV. Note that the size of the arrows here is independent of the $\langle \mathbf{S}_{\parallel} \rangle_{\mathbf{k}}$ here, whilst their colour indicates the sign of $\langle S_{z} \rangle_{\mathbf{k}}$: red (blue) for $\langle S_{z} \rangle_{\mathbf{k}} > 0$ ($< 0$). (d): Angular dependence of the in-plane spin polarisations projections of Fermi surfaces 1 and 2. We omit the spin projection of Fermi surface 3 since it is at most $\sim 0.04$ in magnitude and only non-zero for small disconnected ranges of $\phi$. Clearly the spin projections for A1 possess no parity, whilst for the E1 system they are can be seen to be either odd or even for all Fermi surfaces.}
    \label{Spin_maps_P0}
\end{figure}

We see in the absence of pressure, the MBZ spin projections for the four central bands -- labelled as $n = -2,-1,1,2$ for the second valence, first valence, first conduction, and second conduction band, respectively -- of the symmetrically encapsulated systems are all qualitatively the same, where the in-plane spin projection of the bands, $\langle\mathbf{S}_{\parallel}\rangle_{\mathbf{k}}$, along the $\Gamma_{\text{M}}-M_{\text{M}}$ lines aligns with the momentum, see Fig. \figRef{Spin_maps_P0}{a}. We also note a strong (weak) curving of these spin textures towards the MBZ origin (boundary). In contrast, the asymmetric case with only one layer proximitised exhibits a comparatively unique MBZ spin projection map, where $\langle\mathbf{S}_{\parallel}\rangle_{\mathbf{k}}$ is neither parallel nor perpendicular to the momenta along the $\Gamma_{\text{M}}-M_{\text{M}}$ lines. Additionally, the general inwards/outwards directionality of the spin maps is consistent across all four systems: $n = -2, -1$ host inward pointing spin maps whilst $n = 1, 2$ display outward spin maps. We now turn our attention towards the E1 system for a more direct comparison with the asymmetric system. Here, we find that the encapsulated system displays a larger triangular region of band inversion between the $|n| = 1$ bands compared to the single-layer proximitised case, but has a smaller region of pronounced in-plane spin polarisation. Nonetheless, both exhibit the ability to display the SHE through disorder-driven skew scattering due to having well-defined out-of-plane spin textures, $\langle S_{z} \rangle_{\mathbf{k}} \neq 0$ \cite{Perkins2024}. However, BZ spin maps are not enough to infer the nature of Edelstein response.

The nature of the Edelstein effect can be determined by the spin texture of the Fermi surface. We map out the Fermi surfaces of systems A1 and E1 ($E_{F} = 2$meV and $E_{F} = 3.5$ meV, respectively) by varying the angle of the momentum $\mathbf{k}$ relative to $\Gamma_{\text{M}}$, $\phi$, over fixed values around the MBZ and numerically determine the radial position of the Fermi surface to yield an energy within $\sim 0.1$\% of the Fermi energy. We plot the resulting Fermi surfaces and spin textures in Fig. \figRef{Spin_maps_P0}{b,c}, where we made use of the $D_{3}$ and $C_{3z}$ point group symmetries to relate all portions of the Fermi surface back to the segment of the MBZ with $\phi \in [-\frac{\pi}{3},0)$ for E1 and $\phi \in [-\frac{\pi}{3},\frac{\pi}{3})$ for A1, respectively. The Fermi surfaces appear qualitatively similar, although the spin textures change. As we see, neither system appears to host a radial spin texture, though the spin texture of the E1 system exhibits the expected dihedral symmetry: $\langle S_{x} \rangle_{(k_{x},k_{y})} = \langle S_{x} \rangle_{(k_{x},-k_{y})}$ and $\langle S_{y} \rangle_{(k_{x},k_{y})} = -\langle S_{y} \rangle_{(k_{x},-k_{y})}$ \cite{Perkins2026prep}. To ascertain whether the resulting spin texture permits a collinear Edelstein effect, we project the in-plane spin textures onto the radial and azimuthal unit vectors -- $\langle \mathbf{S}_{\parallel} \rangle_{\text{FS}} \cdot \hat{\mathbf{k}} = \langle S_{x} \rangle_{\text{FS}} \cos \phi + \langle S_{y} \rangle_{\text{FS}} \sin \phi$ and $\langle \mathbf{S}_{\parallel} \rangle_{\text{FS}} \cdot \hat{\boldsymbol{\phi}} = -\langle S_{x} \rangle_{\text{FS}} \sin \phi + \langle S_{y} \rangle_{\text{FS}} \cos \phi$ -- and plot the resulting angular dependence of these projections around the Fermi surface in Fig. \figRef{Spin_maps_P0}{d}. We notice that $\langle \mathbf{S}_{\parallel} \rangle_{\text{FS}} \cdot \hat{\mathbf{k}}$ ($\langle \mathbf{S}_{\parallel} \rangle_{\text{FS}} \cdot \hat{\boldsymbol{\phi}}$) are even (odd) functions of $\phi$ for the E1 system. Therefore, the integrals of these projections around each Fermi surface will be non-zero (vanish), thus indicating a purely CEE is to be expected. Conversely, the A1 system does not exhibit such parities and so hosts neither purely collinear or conventional Edelstein responses.

\begin{figure}[H]
    \centering
    \includegraphics[width=\linewidth]{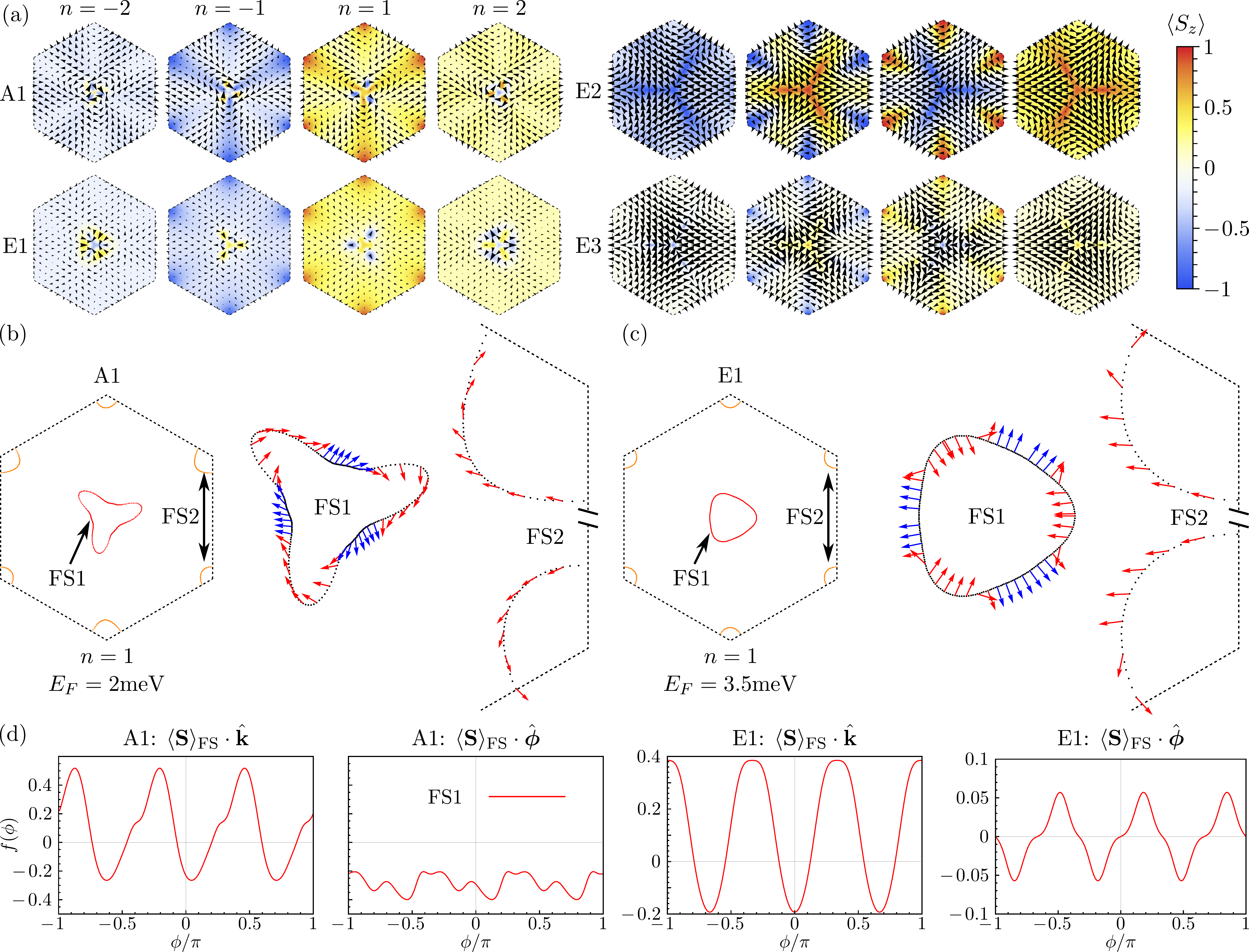}
    \caption{Pressurised spin maps, Fermi surfaces, and spin textures for $P = 1$GPa. (a): Spin maps for the $|n| = 1,2$ bands of the A1, E1, E2, and E3 systems. The arrows indicate the in-plane spin projection at that point in the MBZ, with the arrow size indicating the magnitude of in-plane spin projection: larger arrow heads represent larger values of $\langle \mathbf{S}_{\parallel} \rangle_{\mathbf{k}}$. The colour density plot in the background encodes the out-of-plane spin projection. Band inversions can be seen to occur near $\Gamma_{\text{M}}^{\null}$ ($K_{\text{M}}^{\null}/K_{\text{M}}'$) for A1 and E1 (E2 and E3) in the $|n| = 1$ bands where the sign of $\langle S_{z} \rangle_{\mathbf{k}}$ can be seen to flip relative to the rest of the band. (b)-(c): Fermi surfaces and spin textures for the $n = 1$ band of the A1 system at $E_{F} = 2$ meV (b) and E1 system at $E_{F} = 3.5$ meV (c). Note that the size of the arrows here is independent of the $\langle \mathbf{S}_{\parallel} \rangle_{\mathbf{k}}$ here, whilst their colour indicates the sign of $\langle S_{z} \rangle_{\mathbf{k}}$: red (blue) for $\langle S_{z} \rangle_{\mathbf{k}} > 0$ ($< 0$). (d): Radial and angular projections of the Fermi surface spin texture for the first Fermi surface. We omit the projection of the second Fermi surface due to it being substantially smaller in magnitude.}
    \label{Spin_maps_P1}
\end{figure}

Applying out-of-plane pressure to reduce the layer separation, we map out the spin projections for systems A1 and E1 in Fig. \ref{Spin_maps_P1} and illustrate their spin textures for Fermi energies $E_{F} = 2$ meV and $E_{F} = 3.5$ meV, respectively. By applying such pressure the proximity-induced SOCs will also be enhanced, leading to stronger spin splitting and possible band inversion. For simplicity, we keep the SOC parameters fixed to limit the number of parameters we consider in tuning and explore what enhanced interlayer tunnelling alone enables. We see that the defining characteristic alignment of the spin projection with momentum along the $\Gamma_{\text{M}} - M_{\text{M}}$ for the encapsulated systems survives, although the curving of the spin texture becomes apparent throughout the entire MBZ. Moreover, $\langle \mathbf{S}_{\parallel} \rangle_{\mathbf{k}}$ no longer decays as quickly as the MBZ boundary is approached when compared to the zero pressure case, and appears less heavily concentrated around $\Gamma_{\text{M}}$. Likewise, we see that spin maps for the A1 system at 1GPa of pressure also appear to be more evenly distributed across the entire MBZ, though a clear preference can be seen for larger $\langle \mathbf{S}_{\parallel} \rangle_{\mathbf{k}}$ away from the $\Gamma_{\text{M}}^{\null}-K_{\text{M}}^{\null}$ and $\Gamma_{\text{M}}^{\null}-K_{\text{M}}'$ lines. As in the zero pressure case, the asymmetrically proximitised system does not appear to host simple alignment of the in-plane spin expectation with the momentum along any line in the MBZ.

However, what differentiates the cases of zero pressure and that of 1GPa at the level of $\langle \mathbf{S}_{\parallel} \rangle_{\mathbf{k}}$ is the changing sign of $\langle \mathbf{S}_{\parallel} \rangle_{\mathbf{k}} \cdot \hat{\mathbf{k}}$ across the $\Gamma_{\text{M}}^{\null}-K_{\text{M}}^{\null}$ and $\Gamma_{\text{M}}^{\null}-K_{\text{M}}'$ lines for the A1 and E1 systems. Specifically, at zero pressure, the $\Gamma_{\text{M}}$ point resembles a source or sink with all in-plane spin projections pointing towards or away from the MBZ origin: $\text{sgn}(\langle \mathbf{S}_{\parallel} \rangle_{\mathbf{k}} \cdot \hat{\mathbf{k}})$ appears constant when looping around $\Gamma_{\text{M}}$. However, upon the application of 1GPa of out-of-plane pressure, the spin projection points in opposite radial directions for two neighbouring $M_{\text{M}}$ points, indicating $\text{sgn}(\langle \mathbf{S}_{\parallel} \rangle_{\mathbf{k}} \cdot  \hat{\mathbf{k}})$ changes sign as one circles around the MBZ origin. Interestingly, the E2 and E3 systems exhibited this sign changing behavior for both zero pressure and 1GPa, indicating an intricate interplay of the interlayer tunnelling and SOC energy scales in determining the spin texture. Regarding the out-of-plane spin projection, we see that it changes drastically around $\Gamma_{\text{M}}$ for all systems where new band inversions can be seen to occur between the four central bands, signalling a topological phase transition, see Section \ref{Sec_QG}. As in the zero pressure case, we again see that the encapsulated system will admit a purely collinear Edelstein effect whilst the asymmetric system yields a spin response that is neither perpendicular nor parallel to the applied electrical current, see Figs. \figRef{Spin_maps_P1}{d}.

\section{Topological Tuning} \label{Sec_QG}

\begin{figure}[t]
    \centering
    \includegraphics[width=\linewidth]{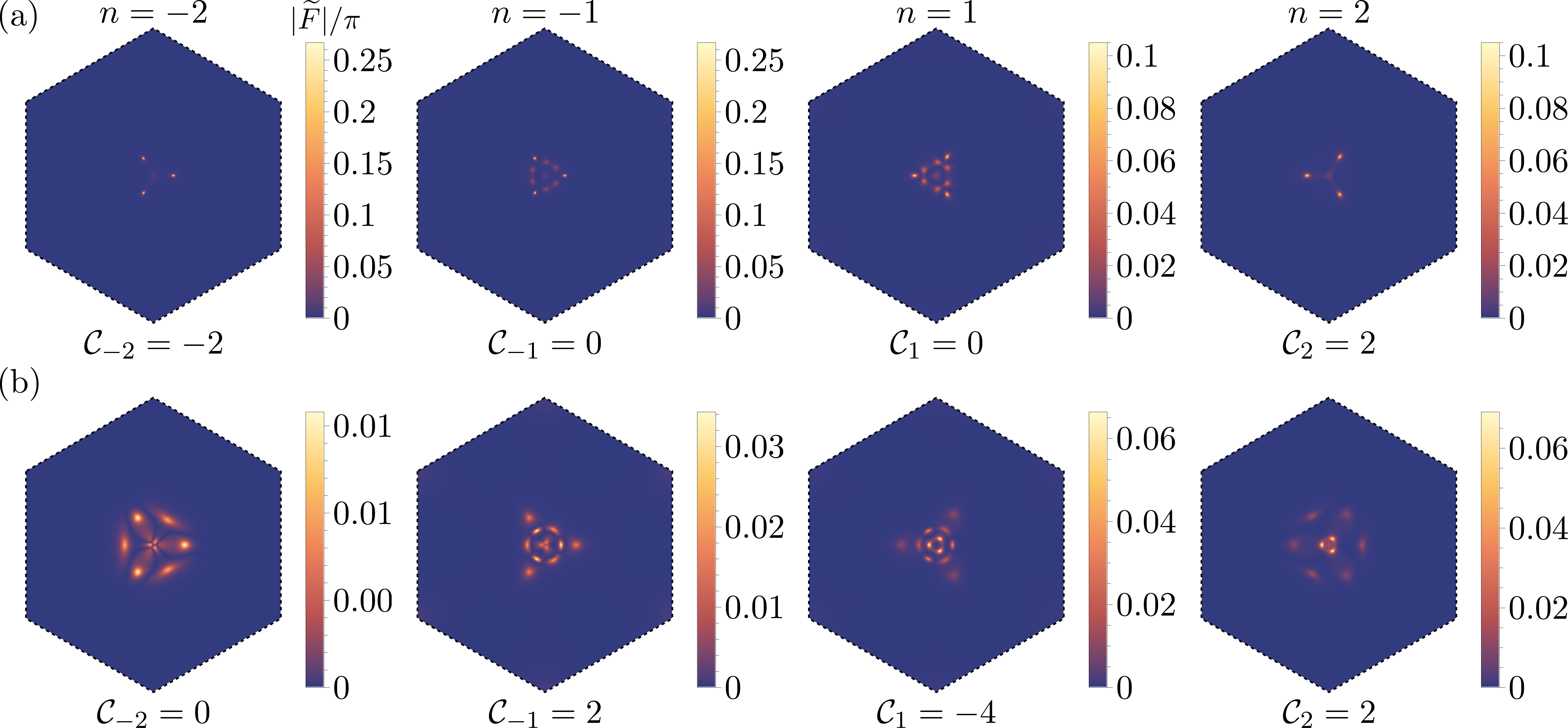}
    \caption{Berry curvature variation depicted by the lattice field, $\widetilde{F}$, as defined by Fukui et al. \cite{Fukui2005} for the spin-split conduction and valence bands at zero pressure (a) and 1 GPa (b) for the E1 system alongside their associated abelian valley Chern number, $\mathcal{C}_{n}$. The lattice field is calculated using a grid of $200 \times 200$ points, giving the relation between Berry curvature and lattice field as $|\Omega| (\delta k)^{2} \approx |\widetilde{F}| < \pi$ with $\delta k = 8\pi \sin(\vartheta)/(199 \sqrt{3})$. The dashed line indicates the MBZ boundary.}
    \label{BC_variation}
\end{figure}

Without applying pressure, the proximity-induced SOCs alone are enough to make the spin-split conduction and valence bands topologically non-trivial as demonstrated in Ref. \cite{Tan2024} by a topological phase diagram and  Fig. \figRef{BC_variation}{a}. We illustrate the distribution of the Berry curvature, $\Omega$, for the E1 system for zero pressure in Fig. \figRef{BC_variation}{a}, where we see it to be highly concentrated around the $\Gamma_{\text{M}}$ point in all four bands, with larger magnitude occurring in the valence bands. From the spin maps of the previous section, we already see that band inversion occurs due to the changing sign of $\angleaverage{S_{z}}$ with applied pressure, indicating a topological phase transition. Whilst this is immediately seen at the level of the valley Chern numbers (Fig. \figRef{BC_variation}{b}), we must also understand how the Berry curvature changes the quantum geometry. Under the application of 1 GPa of out-of-plane pressure, the Berry curvature undergoes a dramatic redistribution between the four bands by spreading out over a wider area. Nonetheless, the quantum geometry is still primarily concentrated towards $\Gamma_{\text{M}}$ but with a more complex landscape of maxima and minima. Moreover, in contrast to the zero-pressure case, the Berry curvature exhibits larger maxima in the conduction bands under this pressure.

\section{Conclusions} \label{Sec_conclusions}

We have shown that encapsulation of MATBG prevents the appearance of a conventional Edelstein effect by only permitting spin textures whose Fermi surface integral is non-zero in the radial direction and vanishing in the azimuthal direction. Ultimately, this will yield a collinear Edelstein effect without the need to fine-tune the twisting angle between the encapsulating TMDs and their neighbouring graphene layers, allowing us to bypass the search for a possible critical angle in proximitised monolayer graphene. Moreover, we demonstrated that out-of-plane pressure can be used to further tune the spin texture of the moir\'{e} system, with the number of active Fermi surfaces changing and the variation of $\angleaverage{\mathbf{S}_{\mathbf{k}}}_{\text{FS}}$ being altered considerably. Nonetheless, the application of pressure did not enable the appearance of a conventional Edelstein effect in any of the encapsulated systems studied here. In addition to tuning the spin texture, we also found pressure to provide another handle to induce topological transitions, resulting in the redistribution of valley Chern numbers and changes in the concentration of quantum geometry. A notable advantage of the pressure tuning is that it may be done in-situ unlike twisting which is fixed once the vdW heterostructure has been synthesised, thus allowing for deeper exploration of the topological phase space of moir\'{e} materials.

We note that for twist angles away from the magic-angle, pressure does not play a similarly significant role since the interlayer tunnelling energies are only slightly altered and remain small compared to the dominant nearest-neighbour intralayer tunnelling energy scale. For example, application of a 1 GPa out-of-plane pressure for $\theta = 1.5^{\circ}$ yields a small shift in the electronic bands but does not yield new band inversions due to the dispersive nature of the bands at zero-pressure. Hence much larger pressures will be required to meaningfully alter the topology and out-of-plane spin texture of the system.

In light of these results, future avenues of interest lie in both spintronics and topology. Regarding the former, a complete map of the spin texture variation as well as the Edelstein and spin Hall responses with both pressure and graphene-TMD twist angles will provide insight on if spin Hall effects or Edelstein effects can be further enhanced beyond those seen in proximity-coupled monolayer graphene. Given we have focused on MATBG, interactions may play a role for weaker SOC where the bands exhibit a small spin splitting and deviation away from their flat nature. However, larger SOC will result in more dispersive bands that may lead to less pronounced interaction effects. Moreover, since pressure can be used as a means to alter the quantum geometry of a material via the Berry curvature, studying the effects of pressure upon other measures of quantum geometry (e.g. Fubini-Study metric \cite{Herzog-Arbeitman2022,Herzog-Arbeitman2024arxiv}) may offer new insight into leveraging quantum materials for technological application.

\section*{Acknowledgements}

DTSP and JJB acknowledge funding from the UK EPSRC through grants EP/X012557/1 and EP/T034351/1. DTSP would also like to thank Aires Ferreira for useful discussions.

\printbibliography

@article{Avsar2020,
  title = {Colloquium: Spintronics in graphene and other two-dimensional materials},
  author = {Avsar, A. and Ochoa, H. and Guinea, F. and \"Ozyilmaz, B. and van Wees, B. J. and Vera-Marun, I. J.},
  journal = {Rev. Mod. Phys.},
  volume = {92},
  issue = {2},
  pages = {021003},
  numpages = {28},
  year = {2020},
  month = {Jun},
  publisher = {American Physical Society},
  doi = {10.1103/RevModPhys.92.021003},
  url = {https://link.aps.org/doi/10.1103/RevModPhys.92.021003},
}

@article{Bernevig2021,
  title = {{Twisted bilayer graphene. I. Matrix elements, approximations, perturbation theory, and a $k\ifmmode\cdot\else\textperiodcentered\fi{}p$ two-band model}},
  author = {Bernevig, B. Andrei and Song, Zhi-Da and Regnault, Nicolas and Lian, Biao},
  journal = {Phys. Rev. B},
  volume = {103},
  issue = {20},
  pages = {205411},
  numpages = {42},
  year = {2021},
  month = {May},
  publisher = {American Physical Society},
  doi = {10.1103/PhysRevB.103.205411},
  url = {https://link.aps.org/doi/10.1103/PhysRevB.103.205411},
}

@article{Bistritzer2011,
  title = {{Moir\'{e} bands in twisted double-layer graphene}},
  author = {Bistritzer, R. and MacDonald, A. H.},
  journal = {Proc. Natl. Acad. Sci. U.S.A.},
  volume = {108},
  issue = {30},
  pages = {12233},
  numpages = {5},
  year = {2011},
  month = {Jul},
  publisher = {National Academy of Sciences},
  doi = {10.1073/pnas.1108174108},
  url = {https://doi.org/10.1073/pnas.1108174108},
}

@article{Cao2018,
  title = {{Unconventional superconductivity in magic-angle graphene superlattices}},
  author = {Cao, Y. and Fatemi, V. and Fang, S. and Watanabe, K. and Taniguchi, T. and Kaxiras, E. and Jarillo-Herrero, P.},
  journal = {Nature},
  volume = {556},
  issue = {7699},
  pages = {43--50},
  year = {2018},
  month = {Mar},
  publisher = {Nature},
  doi = {10.1038/nature26160},
  url = {https://doi.org/10.1038/nature26160},
}

@article{Cao2021,
  title = {{Strange Metal in Magic-Angle Graphene with near Planckian Dissipation}},
  author = {Cao, Y. and Chowdhury, D. and Rodan-Legrain, D. and Rubies-Bigorda, O. and Watanabe, K. and Taniguchi, T. and Senthil, T. and Jarillo-Herrero, P.},
  journal = {Phys. Rev. Lett.},
  volume = {124},
  issue = {7},
  pages = {076801},
  numpages = {7},
  year = {2020},
  month = {Feb},
  publisher = {American Physical Society},
  doi = {10.1103/PhysRevLett.124.076801},
  url = {https://link.aps.org/doi/10.1103/PhysRevLett.124.076801},
}

@article{CastroNeto2009,
  title = {The electronic properties of graphene},
  author = {Castro Neto, A. H. and Guinea, F. and Peres, N. M. R. and Novoselov, K. S. and Geim, A. K.},
  journal = {Rev. Mod. Phys.},
  volume = {81},
  issue = {1},
  pages = {109--162},
  numpages = {0},
  year = {2009},
  month = {Jan},
  publisher = {American Physical Society},
  doi = {10.1103/RevModPhys.81.109},
  url = {https://link.aps.org/doi/10.1103/RevModPhys.81.109}
}

@inbook{Catarina2019,
  author = {Catarina, Gonçalo and Amorim, Bruno and Castro, Eduardo V. and Lopes, João M. V. P. and Lopes, João M. V. P. and Peres, Nuno},
  publisher = {John Wiley \& Sons, Ltd},
  isbn = {9781119468455},
  title = {{Twisted Bilayer Graphene: Low-Energy Physics, Electronic and Optical Properties}},
  booktitle = {Handbook of Graphene Set},
  chapter = {6},
  pages = {177-231},
  year = {2019},
  doi = {https://doi.org/10.1002/9781119468455.ch44},
  url = {https://onlinelibrary.wiley.com/doi/abs/10.1002/9781119468455.ch44},
}

@article{Cuypers2026,
  title = {Evolution of the Berry curvature dipole in uniaxially strained bilayer graphene},
  author = {Cuypers, Karel and Smeyers, Robin and Jorissen, Bert and Covaci, Lucian},
  journal = {SciPost Phys. Core},
  volume = {9},
  pages = {031},
  year = {2026},
  publisher = {SciPost},
  doi = {10.21468/SciPostPhysCore.9.2.031},
  url = {https://scipost.org/10.21468/SciPostPhysCore.9.2.031},
}

@article{Cysne2026,
  title = {Orbital Hall effect from orbital magnetic moments of Bloch states: The role of a new correction term},
  author = {Cysne, Tarik P. and Souza, Ivo and Rappoport, Tatiana G.},
  journal = {Phys. Rev. Res.},
  volume = {8},
  issue = {3},
  pages = {033086},
  numpages = {12},
  year = {2026},
  month = {Jul},
  publisher = {American Physical Society},
  doi = {10.1103/xflp-2y1c},
  url = {https://link.aps.org/doi/10.1103/xflp-2y1c}
}

@article{David2019,
  title = {Induced spin-orbit coupling in twisted graphene--transition metal dichalcogenide heterobilayers: Twistronics meets spintronics},
  author = {David, Alessandro and Rakyta, P\'eter and Korm\'anyos, Andor and Burkard, Guido},
  journal = {Phys. Rev. B},
  volume = {100},
  issue = {8},
  pages = {085412},
  numpages = {15},
  year = {2019},
  month = {Aug},
  publisher = {American Physical Society},
  doi = {10.1103/PhysRevB.100.085412},
  url = {https://link.aps.org/doi/10.1103/PhysRevB.100.085412},
}

@article{Drogler2014,
  title = {{Nanosecond Spin Lifetimes in Single- and Few-Layer Graphene–hBN Heterostructures at Room Temperature}},
  author = {Drögeler, Marc and Volmer, Frank and Wolter, Maik and Terrés, Bernat and Watanabe, Kenji and Taniguchi, Takashi and Güntherodt, Gernot and Stampfer, Christoph and Beschoten, Bernd},
  journal = {Nano Lett.},
  volume = {14},
  number = {11},
  pages = {6050-6055},
  month = {09},
  year = {2014},
  doi = {10.1021/nl501278c},
  url = {https://doi.org/10.1021/nl501278c},
}

@article{Dulisch2025,
  title = {{Electric-Field-Tunable Spin–Orbit Gap in a Bilayer Graphene/WSe${}_{2}$ Quantum Dot}},
  author = {Dulisch, H. and Emmerich, D. and Icking, E. and Hecker, K. and Möller, S. and Müller, L. and Watanabe, K. and Taniguchi, T. and Volk, C. and Stampfer, C.},
  journal = {Nano Lett.},
  volume = {25},
  number = {26},
  pages = {10549-10555},
  year = {2025},
  doi = {10.1021/acs.nanolett.5c02229},
  url = {https://doi.org/10.1021/acs.nanolett.5c02229},
}

@article{Durand2023,
  title = {Optically Active Spin Defects in Few-Layer Thick Hexagonal Boron Nitride},
  author = {Durand, A. and Clua-Provost, T. and Fabre, F. and Kumar, P. and Li, J. and Edgar, J. H. and Udvarhelyi, P. and Gali, A. and Marie, X. and Robert, C. and G\'erard, J. M. and Gil, B. and Cassabois, G. and Jacques, V.},
  journal = {Phys. Rev. Lett.},
  volume = {131},
  issue = {11},
  pages = {116902},
  numpages = {6},
  year = {2023},
  month = {Sep},
  publisher = {American Physical Society},
  doi = {10.1103/PhysRevLett.131.116902},
  url = {https://link.aps.org/doi/10.1103/PhysRevLett.131.116902}
}

@article{Eda2012,
  title = {{Coherent Atomic and Electronic Heterostructures of Single-Layer MoS${}_{2}$}},
  author = {Eda, Goki and Fujita, Takeshi and Yamaguchi, Hisato and Voiry, Damien and Chen, Mingwei and Chhowalla, Manish},
  journal = {ACS Nano},
  volume = {6},
  number = {8},
  pages = {7311-7317},
  year = {2012},
  doi = {10.1021/nn302422x},
  url = {https://doi.org/10.1021/nn302422x},
}

@article{Fang2015,
  title = {{Ab initio tight-binding Hamiltonian for transition metal dichalcogenides}},
  author = {Fang, Shiang and Kuate Defo, Rodrick and Shirodkar, Sharmila N. and Lieu, Simon and Tritsaris, Georgios A. and Kaxiras, Efthimios},
  journal = {Phys. Rev. B},
  volume = {92},
  issue = {20},
  pages = {205108},
  numpages = {15},
  year = {2015},
  month = {Nov},
  publisher = {American Physical Society},
  doi = {10.1103/PhysRevB.92.205108},
  url = {https://link.aps.org/doi/10.1103/PhysRevB.92.205108},
}

@article{Fukui2005,
  author = {Fukui, T. and Hatsugai, Y. and Suzuki, H.},
  title = {{Chern Numbers in Discretized Brillouin Zone: Efficient Method of Computing (Spin) Hall Conductances}},
  journal = {J. Phys. Soc. Jpn},
  volume = {74},
  number = {6},
  pages = {1674-1677},
  year = {2005},
  doi = {10.1143/JPSJ.74.1674},
  URL = {https://doi.org/10.1143/JPSJ.74.1674},
}

@article{Gachter2022,
  title = {Single-Shot Spin Readout in Graphene Quantum Dots},
  author = {G\"achter, Lisa Maria and Garreis, Rebekka and Gerber, Jonas Daniel and Ruckriegel, Max Josef and Tong, Chuyao and Kratochwil, Benedikt and de Vries, Folkert Kornelis and Kurzmann, Annika and Watanabe, Kenji and Taniguchi, Takashi and Ihn, Thomas and Ensslin, Klaus and Huang, Wister Wei},
  journal = {PRX Quantum},
  volume = {3},
  issue = {2},
  pages = {020343},
  numpages = {7},
  year = {2022},
  month = {May},
  publisher = {American Physical Society},
  doi = {10.1103/PRXQuantum.3.020343},
  url = {https://link.aps.org/doi/10.1103/PRXQuantum.3.020343},
}

@article{Gerber2025,
  title = {{Tunable Spin–Orbit Splitting in Bilayer Graphene/WSe${}_{2}$ Quantum Devices}},
  author = {Gerber, Jonas D. and Ersoy, Efe and Masseroni, Michele and Niese, Markus and Laumer, Michael and Denisov, Artem O. and Duprez, Hadrien and Huang, Wister Wei and Adam, Christoph and Ostertag, Lara and Tong, Chuyao and Taniguchi, Takashi and Watanabe, Kenji and Fal’ko, Vladimir I. and Ihn, Thomas and Ensslin, Klaus and Knothe, Angelika},
  journal = {Nano Lett.},
  volume = {25},
  number = {33},
  pages = {12480-12486},
  year = {2025},
  doi = {10.1021/acs.nanolett.5c02309},
  url = {https://doi.org/10.1021/acs.nanolett.5c02309},
}

@article{Go2020,
  title = {Orbital torque: Torque generation by orbital current injection},
  author = {Go, Dongwook and Lee, Hyun-Woo},
  journal = {Phys. Rev. Res.},
  volume = {2},
  issue = {1},
  pages = {013177},
  numpages = {12},
  year = {2020},
  month = {Feb},
  publisher = {American Physical Society},
  doi = {10.1103/PhysRevResearch.2.013177},
  url = {https://link.aps.org/doi/10.1103/PhysRevResearch.2.013177},
}

@article{Herzog-Arbeitman2022,
  title = {Superfluid Weight Bounds from Symmetry and Quantum Geometry in Flat Bands},
  author = {Herzog-Arbeitman, Jonah and Peri, Valerio and Schindler, Frank and Huber, Sebastian D. and Bernevig, B. Andrei},
  journal = {Phys. Rev. Lett.},
  volume = {128},
  issue = {8},
  pages = {087002},
  numpages = {8},
  year = {2022},
  month = {Feb},
  publisher = {American Physical Society},
  doi = {10.1103/PhysRevLett.128.087002},
  url = {https://link.aps.org/doi/10.1103/PhysRevLett.128.087002}
}

@misc{Herzog-Arbeitman2024arxiv,
    author = {Herzog-Arbeitman, J. and Yu, J. and C\u{a}lug\u{a}ru, D. and Hu, H. and Regnault, N. and Liu, C. and Vafek, O. and Coleman, P. and Tsvelik, A. and Song, Z.-D. and Bernevig, B. A.},
    title = {{Topological Heavy Fermion Principle For Flat (Narrow) Bands With Concentrated Quantum Geometry}},
    archivePrefix = {arXiv}, 
    note = {\href{https://doi.org/10.48550/arXiv.2404.07253}{arXiv:2404.07253 [cond-mat.str-el]}},
    year = {2024},
    month = {Oct},
}

@misc{Hu2026arxiv,
    author = {Hu, H. and Shao, Y. and Crippa, L. and C\u{a}lug\u{a}ru, D. and Sangiovanni, G. and Wehling, T. and Glazman, L. I. and Bernevig, B. A.},
    title = {{Twisted Bilayer Graphene Lifetimes At Integer Fillings: An Analytic Result}},
    archivePrefix = {arXiv}, 
    note = {\href{https://doi.org/10.48550/arXiv.2604.14303}{arXiv:2604.14303 [cond-mat.str-el]}},
    year = {2026},
    month = {Apr},
}

@article{Huang2022,
  title = {Carbon and vacancy centers in hexagonal boron nitride},
  author = {Huang, P. and Grzeszczyk, M. and Vaklinova, K. and Watanabe, K. and Taniguchi, T. and Novoselov, K. S. and Koperski, M.},
  journal = {Phys. Rev. B},
  volume = {106},
  issue = {1},
  pages = {014107},
  numpages = {10},
  year = {2022},
  month = {Jul},
  publisher = {American Physical Society},
  doi = {10.1103/PhysRevB.106.014107},
  url = {https://link.aps.org/doi/10.1103/PhysRevB.106.014107}
}

@misc{Hung2026arxiv,
    author = {Hung, Y.-C. and Zhou, X. and Bansil, A.},
    title = {{Breakdown of Topological Inheritance and Twist-Induced Quantum Geometry Reconfiguration in Moir\'{e} Flat Bands}},
    archivePrefix = {arXiv}, 
    note = {\href{https://doi.org/10.48550/arXiv.2603.20849}{arXiv:2603.20849 [cond-mat.mes-hall]}},
    year = {2026},
    month = {Mar},
}

@article{Jaoui2022,
  title = {Quantum critical behaviour in magic-angle twisted bilayer graphene},
  author = {Jaoui, Alexandre and Das, Ipsita and Di Battista, Giorgio and D{\'\i}ez-M{\'e}rida, Jaime and Lu, Xiaobo and Watanabe, Kenji and Taniguchi, Takashi and Ishizuka, Hiroaki and Levitov, Leonid and Efetov, Dmitri K.},
  journal = {Nat. Phys.},
  volume = {18},
  number = {6},
  pages = {633--638},
  year = {2022},
  doi = {10.1038/s41567-022-01556-5},
  url = {https://doi.org/10.1038/s41567-022-01556-5},
}

@article{Jung2015,
  title = {Origin of band gaps in graphene on hexagonal boron nitride},
  author = {Jung, Jeil and DaSilva, Ashley M. and MacDonald, Allan H. and Adam, Shaffique},
  journal = {Nat. Commun.},
  volume = {6},
  number = {1},
  pages = {6308},
  year = {2015},
  doi = {10.1038/ncomms7308},
  url = {https://doi.org/10.1038/ncomms7308},
}

@article{Koshino2018,
  title = {Maximally Localized Wannier Orbitals and the Extended Hubbard Model for Twisted Bilayer Graphene},
  author = {Koshino, Mikito and Yuan, Noah F. Q. and Koretsune, Takashi and Ochi, Masayuki and Kuroki, Kazuhiko and Fu, Liang},
  journal = {Phys. Rev. X},
  volume = {8},
  issue = {3},
  pages = {031087},
  numpages = {12},
  year = {2018},
  month = {Sep},
  publisher = {American Physical Society},
  doi = {10.1103/PhysRevX.8.031087},
  url = {https://link.aps.org/doi/10.1103/PhysRevX.8.031087},
}

@article{Koshino2019,
  title = {Maximally Localized Wannier Orbitals and the Extended Hubbard Model for Twisted Bilayer Graphene},
  author = {Koshino, Mikito and Yuan, Noah F. Q. and Koretsune, Takashi and Ochi, Masayuki and Kuroki, Kazuhiko and Fu, Liang},
  journal = {Phys. Rev. X},
  volume = {8},
  issue = {3},
  pages = {031087},
  numpages = {12},
  year = {2018},
  month = {Sep},
  publisher = {American Physical Society},
  doi = {10.1103/PhysRevX.8.031087},
  url = {https://link.aps.org/doi/10.1103/PhysRevX.8.031087},
}

@article{Ledwith2025,
  title = {Nonlocal Moments and Mott Semimetal in the Chern Bands of Twisted Bilayer Graphene},
  author = {Ledwith, Patrick J. and Dong, Junkai and Vishwanath, Ashvin and Khalaf, Eslam},
  journal = {Phys. Rev. X},
  volume = {15},
  issue = {2},
  pages = {021087},
  numpages = {40},
  year = {2025},
  month = {Jun},
  publisher = {American Physical Society},
  doi = {10.1103/PhysRevX.15.021087},
  url = {https://link.aps.org/doi/10.1103/PhysRevX.15.021087},
}

@article{Li2019,
  title = {Twist-angle dependence of the proximity spin-orbit coupling in graphene on transition-metal dichalcogenides},
  author = {Li, Yang and Koshino, Mikito},
  journal = {Phys. Rev. B},
  volume = {99},
  issue = {7},
  pages = {075438},
  numpages = {9},
  year = {2019},
  month = {Feb},
  publisher = {American Physical Society},
  doi = {10.1103/PhysRevB.99.075438},
  url = {https://link.aps.org/doi/10.1103/PhysRevB.99.075438},
}

@article{Lima2019,
  title = {{Double flat bands in kagome twisted bilayers}},
  author = {Crasto de Lima, F. and Miwa, R. H. and Su\'arez Morell, E.},
  journal = {Phys. Rev. B},
  volume = {100},
  issue = {15},
  pages = {155421},
  numpages = {4},
  year = {2019},
  month = {Oct},
  publisher = {American Physical Society},
  doi = {10.1103/PhysRevB.100.155421},
  url = {https://link.aps.org/doi/10.1103/PhysRevB.100.155421},
}

@article{Liu2024,
  title = {{Quantum geometry in condensed matter}},
  author = {Liu, Tianyu and Qiang, Xiao-Bin and Lu, Hai-Zhou and Xie, X C},
  journal = {Natl. Sci. Rev.},
  volume = {12},
  number = {3},
  pages = {nwae334},
  year = {2025},
  month = {03},
  doi = {10.1093/nsr/nwae334},
  url = {https://doi.org/10.1093/nsr/nwae334},
}

@article{Lopes_dos_Santos2007,
  title = {{Graphene Bilayer with a Twist: Electronic Structure}},
  author = {Lopes dos Santos, J. M. B. and Peres, N. M. R. and Castro Neto, A. H.},
  journal = {Phys. Rev. Lett.},
  volume = {99},
  issue = {25},
  pages = {256802},
  numpages = {4},
  year = {2007},
  month = {Dec},
  publisher = {American Physical Society},
  doi = {10.1103/PhysRevLett.99.256802},
  url = {https://link.aps.org/doi/10.1103/PhysRevLett.99.256802},
}

@article{Lopes_dos_Santos2012,
  title = {Continuum model of the twisted graphene bilayer},
  author = {Lopes dos Santos, J. M. B. and Peres, N. M. R. and Castro Neto, A. H.},
  journal = {Phys. Rev. B},
  volume = {86},
  issue = {15},
  pages = {155449},
  numpages = {12},
  year = {2012},
  month = {Oct},
  publisher = {American Physical Society},
  doi = {10.1103/PhysRevB.86.155449},
  url = {https://link.aps.org/doi/10.1103/PhysRevB.86.155449},
}

@article{Mak2014,
  title = {{The valley Hall effect in MoS${}_{2}$transistors}},
  author = {K. F. Mak  and K. L. McGill  and J. Park  and P. L. McEuen },
  journal = {Science},
  volume = {344},
  number = {6191},
  pages = {1489-1492},
  year = {2014},
  doi = {10.1126/science.1250140},
  url = {https://www.science.org/doi/abs/10.1126/science.1250140},
}

@article{Mak2016,
  title = {{Photonics and optoelectronics of 2D semiconductor transition metal dichalcogenides}},
  author = {Mak, Kin Fai and Shan, Jie},
  journal = {Nat. Photonics},
  volume = {10},
  number = {4},
  pages = {216--226},
  year = {2016},
  doi = {10.1038/nphoton.2015.282},
  url = {https://doi.org/10.1038/nphoton.2015.282},
}

@article{McCann2013,
  title = {The electronic properties of bilayer graphene},
  author = {McCann, E. and Koshino, M.},
  journal = {Rep. Prog. Phys.},
  volume = {76},
  issue = {5},
  pages = {056503},
  year = {2013},
  doi = {10.1088/0034-4885/76/5/056503},
  url = {https://doi.org/10.1088/0034-4885/76/5/056503},
}

@article{Moon2019,
  title = {Quasicrystalline electronic states in ${30}^{\ensuremath{\circ}}$ rotated twisted bilayer graphene},
  author = {Moon, Pilkyung and Koshino, Mikito and Son, Young-Woo},
  journal = {Phys. Rev. B},
  volume = {99},
  issue = {16},
  pages = {165430},
  numpages = {11},
  year = {2019},
  month = {Apr},
  publisher = {American Physical Society},
  doi = {10.1103/PhysRevB.99.165430},
  url = {https://link.aps.org/doi/10.1103/PhysRevB.99.165430},
}

@article{Novoselov2004,
  title = {{Electric Field Effect in Atomically Thin Carbon Films}},
  author = {K. S. Novoselov  and A. K. Geim  and S. V. Morozov  and D. Jiang  and Y. Zhang  and S. V. Dubonos  and I. V. Grigorieva  and A. A. Firsov },
  journal = {Science},
  volume = {306},
  number = {5696},
  pages = {666-669},
  year = {2004},
  doi = {10.1126/science.1102896},
  url = {https://www.science.org/doi/abs/10.1126/science.1102896},
}

@article{Pan2025,
  title = {Topological Valley Transport in Bilayer Graphene Induced by Interlayer Sliding},
  author = {Pan, Jie and Wang, Huanhuan and Zou, Lin and Wang, Xiaoyu and Zhang, Lihao and Dong, Xueyan and Xie, Haibo and Ding, Yi and Zhang, Yuze and Taniguchi, Takashi and Watanabe, Kenji and Wang, Shuxi and Wang, Zhe},
  journal = {Phys. Rev. Lett.},
  volume = {135},
  issue = {12},
  pages = {126603},
  numpages = {7},
  year = {2025},
  month = {Sep},
  publisher = {American Physical Society},
  doi = {10.1103/26q7-dsm1},
  url = {https://link.aps.org/doi/10.1103/26q7-dsm1}
}

@incollection{Perkins2023,
  title = {{Spintronics in 2D graphene-based van der Waals heterostructures}},
  author = {David T.S. Perkins and Aires Ferreira},
  editor = {Tapash Chakraborty},
  booktitle = {Encyclopedia of Condensed Matter Physics (Second Edition)},
  publisher = {Academic Press},
  edition = {Second Edition},
  address = {Oxford},
  pages = {205-222},
  year = {2024},
  isbn = {978-0-323-91408-6},
  doi = {https://doi.org/10.1016/B978-0-323-90800-9.00203-1},
  url = {https://www.sciencedirect.com/science/article/pii/B9780323908009002031},
}

@article{Perkins2024,
  title = {Ultrafast all-electrical universal nanoqubits},
  author = {Perkins, David T. S. and Ferreira, Aires},
  journal = {Phys. Rev. B},
  volume = {109},
  issue = {4},
  pages = {L041411},
  numpages = {7},
  year = {2024},
  month = {Jan},
  publisher = {American Physical Society},
  doi = {10.1103/PhysRevB.109.L041411},
  url = {https://link.aps.org/doi/10.1103/PhysRevB.109.L041411},
}

@article{Perkins2025b,
  title = {{Designing topological high-order Van Hove singularities: Twisted bilayer kagome}},
  author = {Perkins, David T. S. and Chandrasekaran, Anirudh and Betouras, Joseph J.},
  journal = {Phys. Rev. B},
  volume = {112},
  issue = {23},
  pages = {235134},
  numpages = {16},
  year = {2025},
  month = {Dec},
  publisher = {American Physical Society},
  doi = {10.1103/8y2v-kx2w},
  url = {https://link.aps.org/doi/10.1103/8y2v-kx2w},
}

@misc{Perkins2026arxiv,
    author = {Perkins, David T. S. and Betouras, Joseph J.},
    title = {{Twisted Kagome Bilayers: High-Order Van Hove Singularities, Sublattice Interference, Magic Angles, and Possible Topology}},
    archivePrefix = {arXiv}, 
    note = {\href{https://doi.org/10.48550/arXiv.2605.06551}{arXiv:2605.06551 [cond-mat.mes-hall]}},
    year = {2026},
    month = {May},
}

@misc{Perkins2026prep,
    author = {Perkins, David T. S. and Betouras, Joseph J.},
    note = {In preparation.},
}

@article{Pezo2023,
  title = {Orbital Hall physics in two-dimensional Dirac materials},
  author = {Pezo, Armando and Garc\'{\i}a Ovalle, Diego and Manchon, Aur\'elien},
  journal = {Phys. Rev. B},
  volume = {108},
  issue = {7},
  pages = {075427},
  numpages = {10},
  year = {2023},
  month = {Aug},
  publisher = {American Physical Society},
  doi = {10.1103/PhysRevB.108.075427},
  url = {https://link.aps.org/doi/10.1103/PhysRevB.108.075427},
}

@article{Peterfalvi2022,
  title = {Quantum interference tuning of spin-orbit coupling in twisted van der Waals trilayers},
  author = {P\'eterfalvi, Csaba G. and David, Alessandro and Rakyta, P\'eter and Burkard, Guido and Korm\'anyos, Andor},
  journal = {Phys. Rev. Res.},
  volume = {4},
  issue = {2},
  pages = {L022049},
  numpages = {8},
  year = {2022},
  month = {May},
  publisher = {American Physical Society},
  doi = {10.1103/PhysRevResearch.4.L022049},
  url = {https://link.aps.org/doi/10.1103/PhysRevResearch.4.L022049},
}

@article{Qian2014,
  title = {{Quantum spin Hall effect in two-dimensional transition metal dichalcogenides}},
  author = {Xiaofeng Qian  and Junwei Liu  and Liang Fu  and Ju Li },
  journal = {Science},
  volume = {346},
  number = {6215},
  pages = {1344-1347},
  year = {2014},
  doi = {10.1126/science.1256815},
  url = {https://www.science.org/doi/abs/10.1126/science.1256815},
}

@article{Rao2023,
  title = {Ballistic transport spectroscopy of spin-orbit-coupled bands in monolayer graphene on WSe2},
  author = {Rao, Qing and Kang, Wun-Hao and Xue, Hongxia and Ye, Ziqing and Feng, Xuemeng and Watanabe, Kenji and Taniguchi, Takashi and Wang, Ning and Liu, Ming-Hao and Ki, Dong-Keun},
  journal = {Nat. Commun.},
  volume = {14},
  number = {1},
  pages = {6124},
  year = {2023},
  doi = {10.1038/s41467-023-41826-1},
  url = {https://doi.org/10.1038/s41467-023-41826-1},
}

@article{Shi2019,
  title = {{Imaging quantum spin Hall edges in monolayer WTe${}_{2}$}},
  author = {Yanmeng Shi  and Joshua Kahn  and Ben Niu  and Zaiyao Fei  and Bosong Sun  and Xinghan Cai  and Brian A. Francisco  and Di Wu  and Zhi-Xun Shen  and Xiaodong Xu  and David H. Cobden  and Yong-Tao Cui },
  journal = {Sci. Adv.},
  volume = {5},
  number = {2},
  pages = {eaat8799},
  year = {2019},
  doi = {10.1126/sciadv.aat8799},
  url = {https://www.science.org/doi/abs/10.1126/sciadv.aat8799},
}

@article{Sierra2021,
  title = {Van der Waals heterostructures for spintronics and opto-spintronics},
  author = {Sierra, Juan F. and Fabian, Jaroslav and Kawakami, Roland K. and Roche, Stephan and Valenzuela, Sergio O.},
  journal = {Nat. Nanotechnol.},
  volume = {16},
  number = {8},
  pages = {856--868},
  year = {2021},
  doi = {10.1038/s41565-021-00936-x},
  url = {https://doi.org/10.1038/s41565-021-00936-x},
}

@article{Song2019,
  title = {{All Magic Angles in Twisted Bilayer Graphene are Topological}},
  author = {Song, Zhida and Wang, Zhijun and Shi, Wujun and Li, Gang and Fang, Chen and Bernevig, B. Andrei},
  journal = {Phys. Rev. Lett.},
  volume = {123},
  issue = {3},
  pages = {036401},
  numpages = {6},
  year = {2019},
  month = {Jul},
  publisher = {American Physical Society},
  doi = {10.1103/PhysRevLett.123.036401},
  url = {https://link.aps.org/doi/10.1103/PhysRevLett.123.036401},
}

@article{Song2022,
  title = {Magic-Angle Twisted Bilayer Graphene as a Topological Heavy Fermion Problem},
  author = {Song, Zhi-Da and Bernevig, B. Andrei},
  journal = {Phys. Rev. Lett.},
  volume = {129},
  issue = {4},
  pages = {047601},
  numpages = {10},
  year = {2022},
  month = {Jul},
  publisher = {American Physical Society},
  doi = {10.1103/PhysRevLett.129.047601},
  url = {https://link.aps.org/doi/10.1103/PhysRevLett.129.047601},
}

@article{Sousa2022,
    title = {{Weak localisation driven by pseudospin-spin entanglement}},
	author = {Sousa, Frederico and Perkins, David T. S. and Ferreira, Aires},
	journal = {Commun. Phys.},
    volume = {5},
	number = {1},
	pages = {291},
    year = {2022},
	url = {https://doi.org/10.1038/s42005-022-01066-z},
    doi = {10.1038/s42005-022-01066-z},
}

@article{Sousa2020,
  title = {Skew-scattering-induced giant antidamping spin-orbit torques: Collinear and out-of-plane Edelstein effects at two-dimensional material/ferromagnet interfaces},
  author = {Sousa, Frederico and Tatara, Gen and Ferreira, Aires},
  journal = {Phys. Rev. Res.},
  volume = {2},
  issue = {4},
  pages = {043401},
  numpages = {10},
  year = {2020},
  month = {Dec},
  publisher = {American Physical Society},
  doi = {10.1103/PhysRevResearch.2.043401},
  url = {https://link.aps.org/doi/10.1103/PhysRevResearch.2.043401},
}

@article{Su2023,
  title = {{Superconductivity in twisted double bilayer graphene stabilized by WSe${}_{2}$}},
  author = {Su, Ruiheng and Kuiri, Manabendra and Watanabe, Kenji and Taniguchi, Takashi and Folk, Joshua},
  journal = {Nat. Mater.},
  volume = {22},
  number = {11},
  pages = {1332--1337},
  year = {2023},
  doi = {10.1038/s41563-023-01653-7},
  url = {https://doi.org/10.1038/s41563-023-01653-7},
}

@article{Sun2023,
  title = {Determining spin-orbit coupling in graphene by quasiparticle interference imaging},
  author = {Sun, Lihuan and Rademaker, Louk and Mauro, Diego and Scarfato, Alessandro and P{\'a}sztor, {\'A}rp{\'a}d and Guti{\'e}rrez-Lezama, Ignacio and Wang, Zhe and Martinez-Castro, Jose and Morpurgo, Alberto F. and Renner, Christoph},
  journal = {Nat. Commun.},
  volume = {14},
  number = {1},
  pages = {3771},
  year = {2023},
  doi = {10.1038/s41467-023-39453-x},
  url = {https://doi.org/10.1038/s41467-023-39453-x},
}

@article{Tan2024,
  title = {Topological phases, van Hove singularities, and spin texture in magic-angle twisted bilayer graphene in the presence of proximity-induced spin-orbit couplings},
  author = {Tan, Yuting and Chou, Yang-Zhi and Wu, Fengcheng and Das Sarma, Sankar},
  journal = {Phys. Rev. B},
  volume = {110},
  issue = {16},
  pages = {165406},
  numpages = {18},
  year = {2024},
  month = {Oct},
  publisher = {American Physical Society},
  doi = {10.1103/PhysRevB.110.165406},
  url = {https://link.aps.org/doi/10.1103/PhysRevB.110.165406},
}

@article{Tang2017,
  title = {{Quantum spin Hall state in monolayer 1T'-WTe${}_{2}$}},
  author = {Tang, Shujie and Zhang, Chaofan and Wong, Dillon and Pedramrazi, Zahra and Tsai, Hsin-Zon and Jia, Chunjing and Moritz, Brian and Claassen, Martin and Ryu, Hyejin and Kahn, Salman and Jiang, Juan and Yan, Hao and Hashimoto, Makoto and Lu, Donghui and Moore, Robert G. and Hwang, Chan-Cuk and Hwang, Choongyu and Hussain, Zahid and Chen, Yulin and Ugeda, Miguel M. and Liu, Zhi and Xie, Xiaoming and Devereaux, Thomas P. and Crommie, Michael F. and Mo, Sung-Kwan and Shen, Zhi-Xun},
  journal = {Nat. Phys.},
  volume = {13},
  number = {7},
  pages = {683--687},
  year = {2017},
  doi = {10.1038/nphys4174},
  url = {https://doi.org/10.1038/nphys4174},
}

@article{Veneri2022,
  title = {Nonperturbative approach to interfacial spin-orbit torques induced by the Rashba effect},
  author = {Veneri, Alessandro and Perkins, David T. S. and Ferreira, Aires},
  journal = {Phys. Rev. B},
  volume = {106},
  issue = {23},
  pages = {235419},
  numpages = {9},
  year = {2022},
  month = {Dec},
  publisher = {American Physical Society},
  doi = {10.1103/PhysRevB.106.235419},
  url = {https://link.aps.org/doi/10.1103/PhysRevB.106.235419},
}

@article{Veneri2025,
  title = {Extrinsic Orbital Hall Effect: Orbital Skew Scattering and Crossover between Diffusive and Intrinsic Orbital Transport},
  author = {Veneri, Alessandro and Rappoport, Tatiana G. and Ferreira, Aires},
  journal = {Phys. Rev. Lett.},
  volume = {134},
  issue = {13},
  pages = {136201},
  numpages = {6},
  year = {2025},
  month = {Apr},
  publisher = {American Physical Society},
  doi = {10.1103/PhysRevLett.134.136201},
  url = {https://link.aps.org/doi/10.1103/PhysRevLett.134.136201},
}

@misc{Vituri2026arxiv,
    author = {Vituri, Y. and Berg, E.},
    title = {{Controlled Loop Expansion for the Topological Heavy Fermion Model}},
    archivePrefix = {arXiv}, 
    note = {\href{https://doi.org/10.48550/arXiv.2604.14278}{arXiv:2604.14278 [cond-mat.str-el]}},
    year = {2026},
    month = {Apr},
}

@article{Voiry2015,
  title  = {{Phase engineering of transition metal dichalcogenides}},
  author = {Voiry, Damien and Mohite, Aditya and Chhowalla, Manish},
  journal  = {Chem. Soc. Rev.},
  year  = {2015},
  volume  = {44},
  issue  = {9},
  pages  = {2702-2712},
  publisher  = {The Royal Society of Chemistry},
  doi  = {10.1039/C5CS00151J},
  url  = {http://dx.doi.org/10.1039/C5CS00151J},
}

@article{Wang2018,
  title = {{Colloquium: Excitons in atomically thin transition metal dichalcogenides}},
  author = {Wang, Gang and Chernikov, Alexey and Glazov, Mikhail M. and Heinz, Tony F. and Marie, Xavier and Amand, Thierry and Urbaszek, Bernhard},
  journal = {Rev. Mod. Phys.},
  volume = {90},
  issue = {2},
  pages = {021001},
  numpages = {25},
  year = {2018},
  month = {Apr},
  publisher = {American Physical Society},
  doi = {10.1103/RevModPhys.90.021001},
  url = {https://link.aps.org/doi/10.1103/RevModPhys.90.021001},
}

@article{Wang2012,
  title = {{Electronics and optoelectronics of two-dimensional transition metal dichalcogenides}},
  author = {Wang, Qing Hua and Kalantar-Zadeh, Kourosh and Kis, Andras and Coleman, Jonathan N. and Strano, Michael S.},
  journal = {Nat. Nanotechnol.},
  volume = {7},
  number = {11},
  pages = {699--712},
  year = {2012},
  doi = {10.1038/nnano.2012.193},
  url = {https://doi.org/10.1038/nnano.2012.193},
}

@article{Wang2024,
  title = {Pressure-tuned superflat bands and electronic localization in twisted bilayer graphene-like materials},
  author = {Wang, Hongfei and Lei, Dangyuan},
  journal = {J. Appl. Phys.},
  volume = {136},
  number = {4},
  pages = {045107},
  year = {2024},
  month = {Jul},
  doi = {10.1063/5.0207883},
  url = {https://doi.org/10.1063/5.0207883},
}

@article{Wei2024,
  title = {{Linear resistivity at van Hove singularities in twisted bilayer WSe${}_{2}$}},
  author = {LingNan Wei  and Qiaoling Xu  and Yangchen He  and Qingxin Li  and Yan Huang  and Wang Zhu  and Kenji Watanabe  and Takashi Taniguchi  and Martin Claassen  and Daniel A. Rhodes  and Dante M. Kennes  and Lede Xian  and Angel Rubio  and Lei Wang },
  journal = {Proc. Natl. Acad. Sci. U.S.A.},
  volume = {121},
  number = {16},
  pages = {e2321665121},
  year = {2024},
  doi = {10.1073/pnas.2321665121},
  url = {https://www.pnas.org/doi/abs/10.1073/pnas.2321665121},
}

@misc{Wie2026arxiv,
    author = {Wei, N. and von Oppen, F. and Glazman, L. I.},
    title = {{Lifetime and spectral function of topological heavy fermions}},
    archivePrefix = {arXiv}, 
    note = {\href{https://doi.org/10.48550/arXiv.2604.14369}{arXiv:2604.14369 [cond-mat.str-el]}},
    year = {2026},
    month = {Apr},
}

@article{Wu2018,
  title = {{Observation of the quantum spin Hall effect up to 100 kelvin in a monolayer crystal}},
  author = {Sanfeng Wu  and Valla Fatemi  and Quinn D. Gibson  and Kenji Watanabe  and Takashi Taniguchi  and Robert J. Cava  and Pablo Jarillo-Herrero },
  journal = {Science},
  volume = {359},
  number = {6371},
  pages = {76-79},
  year = {2018},
  doi = {10.1126/science.aan6003},
  url = {https://www.science.org/doi/abs/10.1126/science.aan6003},
}

@article{Xia2026,
  title = {Bandwidth-tuned Mott transition and superconductivity in moir{\'e}WSe2},
  author = {Xia, Yiyu and Han, Zhongdong and Zhu, Jiacheng and Zhang, Yichi and Kn{\"u}ppel, Patrick and Watanabe, Kenji and Taniguchi, Takashi and Mak, Kin Fai and Shan, Jie},
  journal = {Nature},
  volume = {650},
  number = {8102},
  pages = {585--591},
  year = {2026},
  doi = {10.1038/s41586-025-10049-3},
  url = {https://doi.org/10.1038/s41586-025-10049-3},
}

@article{Yu2019,
  title = {Dodecagonal bilayer graphene quasicrystal and its approximants},
  author = {Yu, Guodong and Wu, Zewen and Zhan, Zhen and Katsnelson, Mikhail I. and Yuan, Shengjun},
  journal = {npj Comput. Mater.},
  volume = {5},
  number = {1},
  pages = {122},
  year = {2019},
  doi = {10.1038/s41524-019-0258-0},
  url = {https://doi.org/10.1038/s41524-019-0258-0},
}

@article{Yu2025,
  title = {Quantum geometry in quantum materials},
  author = {Yu, Jiabin and Bernevig, B. Andrei and Queiroz, Raquel and Rossi, Enrico and T{\"o}rm{\"a}, P{\"a}ivi and Yang, Bohm-Jung},
  journal = {npj Quantum Mater.},
  volume = {10},
  number = {1},
  pages = {101},
  year = {2025},
  doi = {10.1038/s41535-025-00801-3},
  url = {https://doi.org/10.1038/s41535-025-00801-3},
}

@article{Zhou2021_SC,
  title = {Superconductivity in rhombohedral trilayer graphene},
  author = {Zhou, Haoxin and Xie, Tian and Taniguchi, Takashi and Watanabe, Kenji and Young, Andrea F.},
  journal = {Nature},
  volume = {598},
  number = {7881},
  pages = {434--438},
  year = {2021},
  doi = {10.1038/s41586-021-03926-0},
  url = {https://doi.org/10.1038/s41586-021-03926-0},
}

@article{Zhou2021_qtrMetal,
  title = {Half- and quarter-metals in rhombohedral trilayer graphene},
  author = {Zhou, Haoxin and Xie, Tian and Ghazaryan, Areg and Holder, Tobias and Ehrets, James R. and Spanton, Eric M. and Taniguchi, Takashi and Watanabe, Kenji and Berg, Erez and Serbyn, Maksym and Young, Andrea F.},
  journal = {Nature},
  volume = {598},
  number = {7881},
  pages = {429--433},
  year = {2021},
  doi = {10.1038/s41586-021-03938-w},
  url = {https://doi.org/10.1038/s41586-021-03938-w},
}

@article{Zhou2024,
  title = {{Generation of Isolated Flat Bands with Tunable Numbers through Moir\'{e} Engineering}},
  author = {Zhou, Xiaoting and Hung, Yi-Chun and Wang, Baokai and Bansil, Arun},
  journal = {Phys. Rev. Lett.},
  volume = {133},
  issue = {23},
  pages = {236401},
  numpages = {6},
  year = {2024},
  month = {Dec},
  publisher = {American Physical Society},
  doi = {10.1103/PhysRevLett.133.236401},
  url = {https://link.aps.org/doi/10.1103/PhysRevLett.133.236401},
}

\end{document}